\documentstyle[sprocl]{article}

\input{psfig}
\def\be{\begin{equation}}
\def\ee{\end{equation}}
\def\bea{\begin{eqnarray}}
\def\eea{\end{eqnarray}}

\def\mincir{\ \raise -2.truept\hbox{\rlap{\hbox{$\sim$}}\raise5.truept	
\hbox{$<$}\ }}								%
\def\magcir{\ \raise -2.truept\hbox{\rlap{\hbox{$\sim$}}\raise5.truept	%
\hbox{$>$}\ }}								%

\begin{document}

\title{THE TWOFOLD QUASAR-GALAXY CONNECTION} 
\author{A. Cavaliere, V. Vittorini}

\address {Astrofisica, Dip. Fisica 2a Universit\`a, via Ricerca Scientifica 1, 
00133 Roma}

\maketitle\abstracts 
{The rise and the fall of the QS population 
are explained in terms of massive black holes  
forming/accreting 
during the assemblage of the host galaxies, and of accretion 
rekindled by interactions of the host with companions in a group. 
We compute the LFs out to 
$z= 6$. We also predict the masses of relict black holes 
to be found in many galaxies. We compare the histories of the QSO 
and of the star light.}
  
\section{The rise and fall of the quasars}

The body of  radio, optical and X-ray evidence supporting 
a genuine, if broad peak in the number of bright QSs 
around $z \simeq 3$ has  been reviewed by Shaver et al. 1996. 
The QS population is found to {\it rise} in number 
during the first few Gyrs of the Universe life, down to 
epochs $z \simeq 3$; 
in the QS jargon, this is a negative DE toward increasing $z$.
Later than $z \mincir 3$
the population {\it falls}
in cosmic epoch on a scale of $\simeq 2-3$ Gyr. 
This course has been 
interpreted as dominated by LE, but at least 
two additions are called for (see La Franca \& Cristiani 
1997): at the faint end the LFs grow higher with $z$, and require 
a considerable positive DE; the bright end
of the LFs is flatter at low $z$, with more numerous 
bright objects than previously 
recognized.  

That QSs as {\it individual} sources 
are powered by gas accretion onto a massive black hole has been 
argued many times, and direct evidence is now mounting (see Rees 1997). 
But accretion limited only by radiation pressure 
exponentiates on the short time scale 
$\eta\, t_E \simeq 4\,  10^7$ yr. 
The {\it coherent} rise 
and fall of the QS {\it population} over most of the Universe lifetime 
requires extended coordination. 
We show this to be provided by the hierachical growth of structures, 
first as the host {\it galaxies} are built up, and then 
as {\it groups} form where the hosts interact with companions. 
The transition takes place just at $z \sim 3$; but in 
both dynamical regimes the symmetry of the 
gravitational potential is broken,  
and this provides conditions for 
fueling or for refueling the BHs.

\section{BHs fueled by galaxy formation, and refueled by interactions} 

The hierarchical cosmogony  
envisages the typical density perturbation 
collapsing and virializing at the epoch $t$ to have a mass  
growing  
as $ M_c \propto t^{4/(n+3)}$ when $\Omega =1$; for Cold Dark Matter 
perturbations $n\simeq -2$ applies. 
A similar trend 
is retained when $\Omega_o < 1$ until 
the growth freezes out at $1+z \simeq 1/\Omega_o$.
  
A subgalactic building block of DM with  
$M \sim 10^{9}~ M_{\odot}$  
allows a $10^6~M_{\odot}$ BH to form, involving 
a baryon fraction $\epsilon_* \sim 10^{-3}$ 
(HR 93). The galaxy assemblage goes on 
chaotically with angular momentum {\bf j} $\neq$ const, the 
gas accretion is only Eddington-limited and produces 
luminosities $L\sim L_E \propto M_{BH} \sim \epsilon_*  M$.

The transitional mass from a large {\it galaxy} to a small {\it group} 
is $5\; 10^{12} \;M_{\odot}$. In the hierarchical cosmogony this corresponds 
on average to $z_* \simeq 3 \pm 0.5$, depending on the cosmology and 
on the detailed perturbation model. 
Such values match the range where the bright QSs peak, 
corroborating our view. 

In groups, many 
simulations (see Governato et al. 1996) have shown 
interactions to be frequent and 
effective, due to the high 
density of galaxies $n_g \propto \rho_U (z)$, 
and to the velocity dispersion $V \propto 
M_c ^{(1-n)/12} \sim t$ being still 
close to the galaxian  $v_g$. 
Nearly grazing encounters occur on the time scale $\tau_r 
\sim 1/n_g\; \pi r^2_g \; V$, and produce 
merging (Cavaliere \& Menci 1997) or other strong interactions which  
perturb the potential and again cause {\bf j} $\neq$ const. Indeed, 
aimed simulations (see Barnes \& Hernquist 1991) 
show the gas in both interaction partners to lose {\bf j} and 
be driven down to the main galactic nucleus. 

This view is supported by the 
extensive astronomical evidence showing individual QSs
in hosts either engaged in merging and in current interactions,  
or having close companions even submerged within the galaxian body
(see  Hutchings \& Neff 1992, 
Disney et al. 1995, Bahcall et al. 1996, Hasinger et al. 1996). 
The statistics ranges from a lower limit $\sim 15\%$ 
(Rafanelli et al. 1995) 
up to $\sim 1/2$ of the hosts (Bahcall et al. 1996); the average 
QS environment comprises $\sim 10-20$ galaxies (Fisher et al. 1996).

\section{The evolution of the luminosity function} 

To include such bimodal fueling of the same engines, 
 our LFs (see also CPV 97) comprise at any $L, z$ two components: 
$N(L,z) = N_1(L,z) + N_2(L,z)~$.

$N_1$ accounts for 
BHs newly formed on the scale $t_{dyn}$ of the host buildup  
 a process dominant for $M \mincir  5\, 10^{12}\, M_{\odot}$ and 
$z \magcir 3$. 
It is computed on the basis of the Press \& Schechter  1974 mass function 
$N(M,z)$, using the {\it Eddington-limited} $L\propto M$ 
similarly to 
HR 93, but with two crucial differences: we let 
$M_{BH}$ follow
the hierarchical growth of the host galaxy; we consider the 
probability  $\propto \rho^2_U(z)$ of forming BHs as a prefactor of 
$N(L,~z)$ rather than of $M_{BH} \propto L$. 

$N_2$ comprises BHs  reactivated by interactions, and is dominant in structures 
(especially in groups) 
with $M > 5\, 10^{12}\, M_{\odot}$, at $z\mincir 3$. 
The BHs restart 
their bright career from a low luminosity distributed after $f(L)$;
the reactivation of the $N_r$ dormant BHs occurs 
with probability $\propto 1/ \tau_r$. 
Now the activity is {\it supply-limited}, reaches a top $L_b$ to 
statistically fade off on the time scale $\tau_L \propto L_b/L$. All that is 
expressed by the rate equation 
$$\partial_t N_2+\partial_L(\dot{L} N_2)= f(L)\, N_r/\tau _r-
N_2/ \tau_L~.  \eqno(1)$$
To close the argument, $N_r$ is made up 
by the last term integrated over $L$, and by the newly 
formed BHs described by $N_1$. Thus $N_2$ arises from 
$N_1$ and requires no independent normalization; $N_1$ 
is normalized to the data at $z=4$, and implies today about one relict  
BH per few bright galaxies. 

\begin{figure}
\psfig{figure=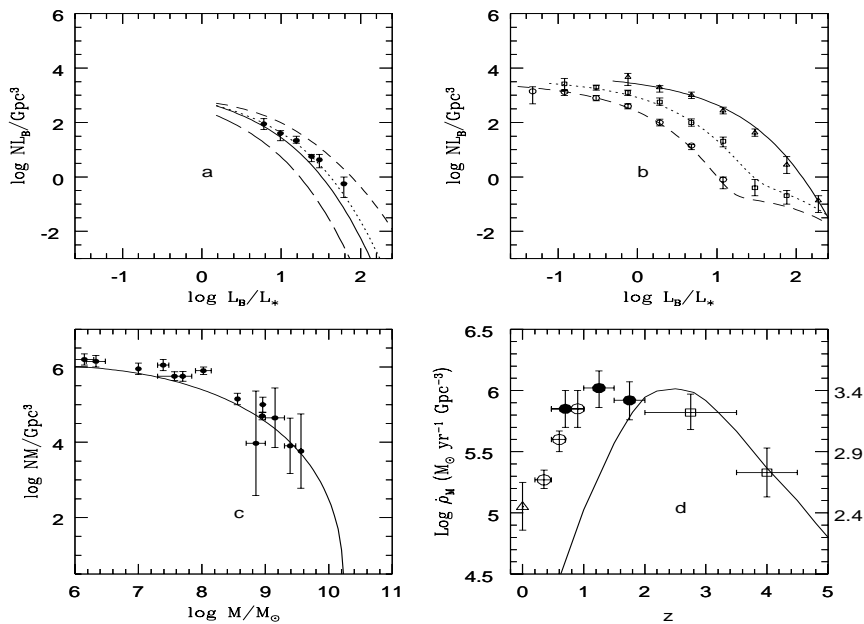,height=3.5in,width=4.7in}
\vspace*{-.5cm} 
\caption{The LFs at $z= 6, 5, 4.5, 3$ (panel {\it a}, from bottom to top),  
and at $z= 2.4, 1, 0.5$ (panel {\it b}, from top to bottom) computed for 
QS forming in the  critical universe from tilted CDM density 
perturbations, and then accreting host gas upon interactions; optically selected 
data as in CPV 97.
Panel {\it c} shows the mass distribution of dormant BHs expected from this model   
at the center of many 
normal galaxies; for the data, 
see CPV 97.
Panel {\it d} compares the history of the light density from QSOs with bolometric 
$L > 10^{45}$ erg/s, to 
that of the star light produced by Madau 1997; the normalization of the former 
is actually $\sim 1/30$. 
\label{fig:fig1}}
\end{figure}
\vspace*{-.2cm}
\section{Conclusions and discussion}

Fig. 1a shows the LFs at high $z$. The prediction at $z =6$ is sensitive 
to the threshold $\delta_c$ for collapse in the Press \& Schechter formula; 
this is taken here to decrease  slowly with $z$ from the value 1.7,  
as indicated by recent simulations. 

Fig. 1b represents the low-$z$ 
behavior when the accreted gas is provided mainly by the {\it host} reservoir 
(say, a constant fraction used up in each interaction); 
then $-\dot M_{gas}/M_{gas} 
\simeq \tau_r^{-1} \simeq -\dot L_b/L_b$ holds, to yield 
$L_b(t) \propto t^ {-t_o/\tau_{ro}}$.
With $\tau_{ro} \simeq 6$ Gyr (scaled to groups 
from the classic census of local interacting galaxies by Toomre 1977), the result 
is 
$L_b \propto (1+z)^{3}$. This implies LE dominant at $z < 3$, 
and LFs flattened by L increasing over the flyby time $2r_g/V \sim$
few $10^8$ yr. 
But fig. 1b shows that quite some DE also 
occurs, since the frequency of the effective interactions peters off 
when groups aggregate into clusters with high $V$. 
Fig. 1c shows the mass distribution predicted for the relict BHs in many 
local, currently inactive galactic nuclei.

Alternatively, the gas may reach the nucleus of a large galaxy 
mainly from {\it satellites} cannibalized out of an initial 
retinue gradually depleted. 
Then  $L_b \sim$ const obtains with no LE, while the DE is enhanced; 
the gas mass $\propto M_{sat}$ follows the satellite distribution and 
yields steeper LFs at the faint end.  

In either case, one ends up with the {\it QS light} history (of gravitational
origin)
shown in fig. 1d, and compared with the {\it star light} history (of 
thermonuclear origin) produced by Madau 1997; at $z\mincir 1$ this is  
mainly contributed by 
the faint blue galaxies, which 
 Cavaliere \& Menci 1997 interpret as starbursts in dwarfs interacting in 
LSS. The similar run 
at such $z$ corroborates the view 
that interactions in dense environments (condensing LSS or virialized 
groups, respectively) drive much of the light illuminating the 
$z\sim 1$ Universe.

To appear in Proc. 12th Potsdam Cosmology Workshop, Sept. 1997, 

``Large-Scale Structures: Tracks and Traces'', 

eds. V. M\"uller et al., 

World Scientific, Singapore

\end{document}